# Realization of the kagome spin ice state in a frustrated intermetallic compound

Kan Zhao[1*+], Hao Deng[2+], Hua Chen[3+], Kate A. Ross[3], Vaclav Petříček[4], Gerrit Günther[5], Margarita Russina[5], Vladimir Hutanu[2], and Philipp Gegenwart[1*]

[1] Experimentalphysik VI, Center for Electronic Correlations and Magnetism, University of Augsburg, 86159 Augsburg, Germany.

[2] Institute of Crystallography, RWTH Aachen University and Jülich Centre for Neutron Science (JCNS) at Heinz Maier-Leibnitz Zentrum (MLZ), Garching, Germany.

[3] Department of Physics, Colorado State University, Fort Collins, CO 80523-1875, USA.

[4] Institute of Physics, Academy of Sciences of the Czech Republic, Na Slovance 2, 18221 Prague, Czech Republic.

[5] Helmholtz-Zentrum Berlin für Materialien und Energie, D-14109 Berlin, Germany

+These authors contributed equally: Kan Zhao, Hao Deng, and Hua Chen

*Corresponding author. Email: kan.zhao@physik.uni-augsburg.de, and philipp.gegenwart@physik.uni-augsburg.de

**Abstract:** Spin ices are exotic phases of matter characterized by frustrated spins obeying local "ice rules", in analogy with the electric dipoles in water ice. In two dimensions, one can similarly define ice rules for in-plane Ising-like spins arranged on a kagome lattice. These ice rules require each triangle plaquette to have a single monopole, and can lead to various unique orders and excitations. Using experimental and theoretical approaches including magnetometry, thermodynamic measurements, neutron scattering and Monte Carlo simulations, we establish HoAgGe as a crystalline (i.e. non-artificial) system that realizes the kagome spin ice state. The system features a variety of partially and fully ordered states and a sequence of field-induced phases at low temperatures, all consistent with the kagome ice rule.

Frustration in spin systems can result in the formation of exotic phases of matter (*1*). One example is the pyrochlore spin ice, in which four nearest-neighbor Ising-like spins sitting at the vertices of a tetrahedron are forced by the exchange and dipolar interactions to obey the "ice rule": two spins pointing into and the other two pointing out of the tetrahedron. Such a local constraint can lead to a macroscopic number of degenerate ground states or an extensive ground state entropy (*2-8*).

In two dimensions (2D), ice rules can be similarly defined for in-plane Ising like classical spins residing on the kagome lattice (*9-11*), which require two-in-one-out or one-in-two-out local arrangements of the spins on its triangles. By viewing each spin effectively as a magnetic dipole formed by two opposite magnetic charges or monopoles, the ice rule leaves either a positive or a negative monopole ($Q_m=\pm 1$) at each triangle, and gives a ground state entropy of ~$0.501 k_B$ per spin, where $k_B$ is the Boltzmann constant. However, a $\sqrt{3} \times \sqrt{3}$ ground state can be selected by further neighbour exchange couplings or the long-range dipolar interaction (*9-11*). Consequently, kagome spin ices show a characteristic multi-stage ordering behavior under changing temperature.

Experimentally, kagome spin ices have only been realized in artificial spin ice systems formed by nanorods of ferromagnets organized into honeycomb networks (*12-18*). However, the large magnetic energy scales and system sizes make it challenging to explore the rich phase diagram of spin ices in the thermodynamic limit (*17, 18*). Alternatively, kagome ice behavior has been reported in pyrochlore spin ices such as $Dy_2Ti_2O_7$ and $Ho_2Ti_2O_7$ under magnetic field along the [111] direction (*19-21*). At the right strength, such a magnetic field can align the Ising spins on the triangular layers of the pyrochlore structure; because the field does not break the ice rule, the in-plane components of the spins on the kagome layers can satisfy the kagome ice rule. However, this is true only in a narrow range of field strength (less than 1T), owing to the weak exchange/dipolar interactions in such systems. Most recently, a magnetic charge order has been suggested in the tripod kagome compound $Dy_3Mg_2Sb_3O_{14}$(*22, 23*), and a dynamic kagome ice has been observed in $Nd_2Zr_2O_7$ under field along the [111] direction (*24, 25*), but a long-range spin order does not appear in either case even at the lowest temperature.

Here, we use multiple experimental and theoretical approaches to show that the intermetallic compound HoAgGe is a naturally existing kagome spin ice that exhibits a fully ordered ground state.

**Structure and magnetometry measurements**

HoAgGe is one of the ZrNiAl-type intermetallics with space group P-62m, which is non-centrosymmetric. In particular, Zr sites in the *ab* plane form a distorted kagome lattice (*26, 27*) [Fig. 1A]. The distortion is characterized by opposite rotations of the two types of triangles in the kagome lattice by the same angle (~15.58º in HoAgGe) around the *c* axis. The rotation breaks the spatial inversion symmetry of a single kagome layer, although it does not change the space group of the 3D crystal (*28*). Previous neutron diffraction measurements suggested the presence of noncollinear magnetic structures of HoAgGe (*29*), but the powder samples used in that work yielded limited magnetic peaks that were insufficient to fully determine the magnetic structure, especially in the presence of frustration. Below we combine neutron diffraction with thermodynamic measurements in single-crystalline HoAgGe to reveal its exotic temperature- and magnetic-field-dependent magnetic structures, which we show to be consistent with the kagome ice rule.

Each $Ho^{3+}$ atom in HoAgGe has ten 4*f* electrons. According to Hund's rules they should have the ground state of $^5I_8$ with an effective magnetic moment $\mu_{eff}$=10.6 $\mu_B$, as confirmed by our Curie-Weiss fitting to the anisotropic inverse susceptibilities $\chi^{-1}(T)$ above 100 K [Fig. S2A]. At lower temperatures, $\chi(T)$ for **H**//*b* under 500 Oe exhibits a relatively sharp peak at 11.6 K (denoted as $T_2$) and another broad inflection at ~7 K (denoted as $T_1$), which are more clearly seen in the plot of the temperature derivative of $\chi(T)$ [Fig. 1B]. Similar behaviors are also observed for **H**//*a*, whereas for **H**//*c* $\chi(T)$ monotonically increases with decreasing temperature (*27*).

More interestingly, plots of magnetization vs. **H**//*b* show a series of plateaus at low temperatures [Fig. 1C]. At *T*=5 K one can clearly identify three metamagnetic transitions at $H \approx$ 1T, 2T, and 3.5T. At each transition the magnetization changes roughly by 1/3 of the saturated value ($M_s$) at *H*>4T. At lower temperatures [1.8 K in Fig. 1C], two additional small plateaus with a jump of ~1/6 $M_s$ appear at 0.9T and 3.2T, respectively, accompanied by a small hysteresis. *M*(*H*) curve for **H**//*a* also shows well defined plateaus, albeit at different ranges of field [Fig. S2B], whereas no plateaus are observed for **H**//*c* [Fig. S2E]. An *H-T* phase diagram based on temperature dependence of the peaks in field derivative of *M*(*H*) curves [Fig. S2C-D] is constructed in Fig. 1D. Together with the $\chi(T)$ data above, the lack of any clear magnetic transitions for **H**//*c* confirms that the Ho spins in HoAgGe are constrained in the *ab* plane, and have additional in-plane anisotropies, similar to that in the isostructural TmAgGe and TbPtIn (*30*).

**Magnetic structures determined from neutron diffraction**

To fully determine the nontrivial spin structures of HoAgGe, we performed single-crystal neutron diffraction experiments down to 1.8K and under **H**//*b* up to 4T (*31*). Below the high-temperature transition $T_2$=11.6K, a magnetic peak appears at (1/3, 1/3, 0) (Fig. 2A and Fig. S4A), indicating a $\sqrt{3} \times \sqrt{3}$ magnetic unit cell [the green rhombus in Fig. 2B]. Below 10K, most nuclear sites exhibit almost constant intensity, but the broad transition at $T_1$ induces additional magnetic contribution at certain structural diffraction sites, such as (1, 0, 0) [Fig. 2A, inset].

According to neutron data at 10K (Fig. S5A), the magnetic structure belongs to the magnetic space group P-6'm2' (Table 1), which has three nonequivalent Ho sites labeled by Ho1, Ho2, and Ho3 in Fig. 2B. Six other Ho positions in the magnetic unit cell are obtained from above three by three-fold rotations around the *c* axis. Because there are no magnetic contributions at nuclear sites at 10K, the simplest possibility for ($M_{Ho1}$, $M_{Ho2}$, $M_{Ho3}$) is ($M$, -$M$, 0), with $M$ determined to be 5.2(1)$\mu_B$ (Table 1 and Fig. S10B). This corresponds to Ho1, Ho2 exhibiting ordered moments of the same size but opposite directions, and Ho3's moment fluctuating without ordering. Such a partially-ordered magnetic structure is shown in Fig. 2C, with the ordered moments forming clockwise or counterclockwise hexagons separated by the unordered moments. The structure thus has a nonzero magnetic toroidal moment defined by $\boldsymbol{\tau} = \frac{1}{V}\int d^3 r\, \mathbf{r} \times \mathbf{M}$ (*32*). Similar partially-ordered structures have also been observed in the isostructural Kondo lattice CePdAl below 2.7K with easy *c*-axis anisotropy (*33*), and in hexagonal UNi$_4$B below 20K with two thirds of U moments forming in plane clockwise hexagons (*34*).

Below $T_1$~7 K, Ho3 moments also enter the long-range order, as indicated in the inset of Fig. 2A. Refinement of neutron data at 4K (Fig. S5B and Fig. S10C) leads to the magnetic structure shown in Fig. 2E, which also has the P-6'm2' symmetry, with ($M_{Ho1}$, $M_{Ho2}$, $M_{Ho3}$)=($M$, -$M$, -$M$) and $M$ =7.5(1) $\mu_B$. As illustrated in Fig. 2F, this fully ordered ground state includes alternating clockwise and counterclockwise hexagons of spins, and another 1/3 of hexagons consisting of three pairs of parallel spins. This is exactly the $\sqrt{3} \times \sqrt{3}$ ground state of the classical kagome spin ice predicted theoretically before (*35-37*).

To confirm that HoAgGe is indeed a kagome spin ice, however, it is necessary to show that the ice rule is established even outside the fully ordered ground state (*9-11*). The kagome ice rule requires dominating nearest-neighbor ferromagnetic coupling between coplanar spins with site-dependent Ising-like uniaxial anisotropy (*9-11*). Using neutron diffraction under magnetic fields we show that these requisites are indeed satisfied in HoAgGe. Figure 2D displays the neutron-scattering integrated intensities of the magnetic peaks at (-1/3, 2/3, 1) (Fig. S4B) and (1/3, 4/3, 1) vs. the strength of the magnetic field along the *b* axis at 4K. Overall the intensity decreases with increasing field and disappears at *H*>3.2T, with sudden changes at the metamagnetic transitions depicted in Figs. 1, C and D, suggesting the shrinking of the magnetic unit cell in field. To obtain further information, we refine the magnetic structures at the three major *M*(*H*) plateaus from the neutron scattering. The magnetic field breaks the 3-fold rotational symmetry and turns the ground state magnetic space group P-6'm2' into Am'm2', with the 9 Ho moments in the $\sqrt{3} \times \sqrt{3}$ unit cell forming 6 nonequivalent groups [Fig. 2G].

Figures 2, G-I, show the magnetic structures at the three major plateaus, obtained from the neutron data taken at 1.8K and *H*=1.5T, 2.5T, and 4T along the *b* axis, respectively (also see Table 1 for the refinement factors). One first notices that all of them can be obtained from the ground state by reversing certain Ho spins, with negligible rotation from their local Ising axis (*31*). This is strong evidence for the Ising-like anisotropy of the Ho moments, with the local easy axes defined by a perpendicular mirror plane through each atom. The Ising-like anisotropy is further confirmed by our Crystalline Electric Field (CEF) calculations below. Moreover, in all three structures the spins are always reversed in such a way that the one-in-two-out or two-in-one-out ice rule is satisfied, but the total magnetic moment along *b* increases with increasing field. At *H*=4T, the magnetic unit cell becomes identical to the structural unit cell (*14, 15, 18*), and has the largest possible net moment allowed by the ice rule. This is further corroborated by the identical magnetization jump of 1.7$\mu_B$/Ho at the three metamagnetic transitions at 1.8K (Fig. S2B). Assuming the magnetic structures in Figs. 2, G-I, this jump can be translated to an ordered moment size of *M*=(9/2)×1.7$\mu_B$=7.65$\mu_B$, roughly consistent with that determined from neutron data at zero field [7.5(1)$\mu_B$ at 4K and 5.2(1)$\mu_B$ at 10K]. These results indicate that the Ho moments at low temperatures are constrained by the kagome ice rule. The metamagnetic

transitions result from the competition between the external magnetic field and the weaker, further than nearest-neighbor couplings that do not affect the ice rule. For a detailed analysis of the three magnetic structures, see (*31*).

**Specific heat and magnetic entropy**

Having established the existence of the kagome ice rule in HoAgGe at low temperatures, we now proceed to examine the thermodynamic behaviors of kagome spin ice. To this end we isolate the magnetic contribution to the specific heat $C_{mag}$ by subtracting the contributions from nuclei, lattice vibrations, and itinerant electrons (*31*). Figure 3A shows the $C_{mag}$ thus obtained from 136K down to 0.48K. Besides the two peaks at $T_1$ and $T_2$, another broad peak appears at 26K that is discussed further below.

Figure 3B shows the magnetic entropy $S_m(T)$ obtained by integrating $C_{mag}(T)/T$ from (nominally) $T$=0K. At high temperatures >100K, $S_m$ approaches Rln17, consistent with the $^5I_8$ state of an isolated $Ho^{3+}$ and close to that of the structurally similar intermetallic compounds $HoNiGe_3$ (*38*) and $Ho_3Ru_4Al_{12}$ (*39*). For the ideal kagome spin ice, however, $S_m$ should approach Rln2 at high temperatures because of the Ising anisotropy. The temperature dependence of the magnetic entropy of HoAgGe thus must be analyzed together with the CEF splitting of the $Ho^{3+}$ $J$=8 multiplet (see below).

Short-range spin-ice correlations stemming from kagome ice rule can lead to a broad peak in specific heat $C_{mag}(T)$ at the temperature scale corresponding to the nearest neighbor exchange coupling (*10, 11*). To investigate the origin of the broad peak at 26 K in Fig. 3A, we also investigated $Lu_{1-x}Ho_xAgGe$ ($x$=0.52 and 0.73). Because $Lu^{3+}$ is not magnetic, the exchange interaction between Ho moments is suppressed as $x$ decreases, whereas the CEF splitting that can lead to the Schottky anomaly should not change much. As shown in Fig. S11, the $T_1$ and $T_2$ for the magnetic transitions shift down to 8K and 4K for $Lu_{0.27}Ho_{0.73}AgGe$, and for

Lu$_{0.48}$Ho$_{0.52}$AgGe $T_2$ shifts to 5 K with $T_1$ <1.8K. However, in both cases the broad anomaly in the $C_{mag}$ curves still appears at about 26K. We therefore conclude that the broad peak is a Schottky anomaly caused by the CEF splitting of Ho$^{3+}$ multiplet.

To clearly see the effects of short-range correlations caused by the exchange interaction between Ho moments, we subtract the normalized Lu$_{1-x}$Ho$_x$AgGe magnetic specific heat from that of pure HoAgGe [Fig. 3C]. The resulting $\Delta C_{mag}$ is almost constant (within the error bar) above 20 K, but increases as $T$ goes below 20K until reaching a maximum at the transition to the partially ordered state; therefore short-range spin ice correlations still exist below 20K. However, the broad peak characteristic of an ideal kagome ice model (*10, 11*) is absent, which will be discussed further below.

**Inelastic neutron scattering and CEF analysis**

To see to what extent the Ho spins can be approximately viewed as Ising, we next discuss the CEF effects. According to the local orthorhombic symmetry (point group $C_{2v}$) of Ho sites in HoAgGe, CEF splits the 17-fold multiplet of a non-Kramers Ho$^{3+}$ ion into 17 singlets. To directly probe the CEF splitting, we conducted inelastic neutron scattering (INS) experiments of HoAgGe crystals using time-of-flight (TOF) spectrometer NEAT at Helmholtz Zentrum Berlin (*31, 40*). To minimize the influence of internal fields caused by magnetic exchange interactions in the presence of long range order (*41*), we choose to conduct the measurements at 10K, very close to $T_2$, with incident neutron wavelengths 2.4Å and 3Å. Clear CEF modes, which are independent of momentum transfer Q, appear between 4 meV and 6 meV [Fig. 3D]. The same modes are also observed in the INS spectra at 50K in Fig. S14B indicating the internal field is already very weak at 10K. Additional CEF modes with broad features appear between 8 meV and 11 meV in Fig. S14A. With incident neutron wavelength 5 Å (3.27 meV) at 15K, a continuum feature appears in the quasi-elastic scattering plane ($\Delta E$<0.2 meV) in Fig. S14C, indicating diffuse scattering, which is consistent with the observation of short-range spin correlations below 20K in Fig. 3C.

Based on the INS data at 10K and the magnetic specific heat result above 20K in Fig. 3A, we did a combined fitting [Fig. 3E] to obtain the CEF Hamiltonian. The nine CEF parameters from the fitting are listed in Table S3. As shown in Table 2, among the 17 CEF levels, the lowest four with energies less than 1 meV should be the major ones contributing to the kagome ice behavior at low temperatures. The other 13 CEF modes are listed in Table S4. The 4 low energy CEF modes indeed have Ising-type anisotropy as shown in Fig. 3F. Under a magnetic field along the local Ising axis, the Ho moment steeply increases to 7.7 $\mu_B$ at 1T (8.1$\mu_B$ at 6T), which is much larger than 6.5$\mu_B$ and 4.0$\mu_B$ for fields along the two perpendicular directions at 6T. Because the CEF Hamiltonian does not include the effect of exchange coupling between Ho moments, which is of similar size to the separation between the lower CEF levels, it may not fully account for the anisotropy of the moments, especially at low temperatures.

**Classical Monte Carlo simulations**

Based on the experimental evidence presented above, we propose a classical spin model consisting of Ising-like in-plane spins on the 2D distorted kagome lattice of the [0001] plane of HoAgGe. The spins are coupled to one another through exchange couplings and long-range dipolar interactions and to external magnetic fields through Zeeman coupling. The comprehensive *M*(*H*) data and magnetic structures from neutron scattering allow us to extract the exchange couplings up to the 4th nearest neighbor, with implicit summation over periodic images along the *c* axis (*31*). The exchange couplings are found to be dominant over the dipolar interaction, quite different from the pyrochlore spin ices $Dy_2Ti_2O_7$ and $Ho_2Ti_2O_7$, as well as $Dy_3Mg_2Sb_3O_{14}$ (*22, 23*) where they are comparable. This is likely a result of the relatively strong RKKY-type interaction between Ho moments (*27*). As a check we calculated the *M*(*H*) curves for **H** along *b* and *a* axes at *T*=1K through Monte Carlo simulations [Fig. 4A], which agree with the experimental results in Fig. 1C and Fig. S2B. We do not consider the effects of the Dzyaloshinkiy-Moriya interaction (*31*).

Monte Carlo simulations of the classical spin model on an 18×18 lattice with periodic boundary conditions give three peaks in the specific heat vs. temperature plot as in the experiment [Fig. 4B]. The positions of the two peaks of $C(T)$ at lower temperatures agree well with the experimental data in Fig. 3B. The broad peak at the highest temperature (denoted as $T_3$) is due to the gradual development of ice-rule correlations for Ising spins. $T_3$ is mainly determined by the nearest neighbor exchange coupling which, however, is not fixed by the experimental data. Because $Ho^{3+}$ in the real material is not Ising-like at $T>20$ K thanks to strong population of higher CEF levels, such a peak does not have to be present in experiment (*31*). Most importantly, the ground state and the partially-ordered state can both be reproduced by the Monte Carlo simulations (Fig. S16). We thus believe that the classical spin model captures the main characteristics of the magnetism of HoAgGe at low temperatures.

The temperature dependence of the magnetic entropy in Fig. 4C confirms that the peaks of $C(T)$ at $T_3$ and $T_1$ in Fig. 4B correspond to the formation of spin-ice correlations and the fully-ordered ground state, respectively, similar as predicted for dipolar kagome ice (*10, 11*). However, the peak at $T_2$ in Fig. 4B does not correspond to the transition into the "magnetic charge order" in dipolar kagome ice, which has an entropy of 0.108 $k_B$ per spin (*10, 11*). In fact, the magnetic charge order is destabilized by the further neighbor exchange couplings in our model. In contrast, the Monte Carlo simulations of the short-range kagome ice, with ferromagnetic nearest neighbor coupling and antiferromagnetic 2nd nearest neighbor coupling (*9*) indeed give an intermediate state similar to that in Fig. 2C through 1st-order transitions. Simple counting gives an entropy of $\frac{k_B}{3}\ln 2 \approx 0.231\ k_B$ per spin for the partially ordered state shown in Fig. 2B. However, it has been shown more recently that the state of Fig. 2C (and the ground state as well) has a $Z_6$ order parameter, similar to a 6-state clock model, and the transition into it should be of Kosterlitz-Thouless (KT) type (*37*).

Our model (as well as the physical system) is different from both dipolar and short-range kagome ice cases because of the co-existence of the further neighbor exchange couplings and the long-range dipolar interaction. The precise nature of the two low-temperature transitions in our 2D model may only be elucidated through comprehensive finite-size scaling analysis, which deserves a separate study. It also remains an open question how the transitions are influenced by the long-range tail of the RKKY interaction. In reality, however, the KT transition may not be very likely in the 3D HoAgGe, especially considering the strong exchange coupling between neighboring kagome planes (*31*). Another piece of experimental evidence is the critical exponent $\beta=0.321(3)$ of the (1/3, 1/3, 0) peak intensity near $T_2$ [Fig. S4A], which indicates the 3D Ising nature of the magnetic order (*42*).

**Discussion**

Although the Monte Carlo simulations of the classical spin model above are in partial agreement with our experiments, they do not explain the experimental value of the magnetic entropy $S_m$=10.38 J·mol$^{-1}$K$^{-1}$≈1.248$R$ at $T_2$, which is very different from the 0.231$R$ given by the model. Qualitatively the discrepancy should be a result of the thermal population of multiple low-lying CEF levels of Ho$^{3+}$, which, however, leads to the question why the classical Ising Hamiltonian is applicable. In pyrochlore systems such as Dy$_2$Ti$_2$O$_7$, the classical Ising behavior is a consequence of the dominance of CEF splitting over exchange and dipolar interactions. This leads to an effective pseudospin-1/2 Hamiltonian with only the pseudospin components along the local easy axes present (*43*). These pseudospin components are thus good quantum numbers, justifying the use of the classical Ising Hamiltonian.

In HoAgGe, metallicity simultaneously suppresses the CEF splitting of Ho$^{3+}$ ions and enhances the exchange coupling between them, making the two energy scales comparable at least for the low-lying CEF levels. Thus the large ($J$=8) $Ho^{3+}$ moments in HoAgGe at moderately low temperatures, when multiple CEF levels are occupied, are closer to semiclassical spins with strong single-ion anisotropy. Such a semiclassical model can still be mapped to an Ising model at the expense of introducing further neighbor exchange interactions (*44*), which serves as an explanation for the apparent validity of the classical Ising Hamiltonian for HoAgGe. A complete understanding of the entropy data awaits a full quantum mechanical description of the system. It is worth noting that the deviation from an ideal spin-1/2 system can also lead to stronger quantum fluctuations as in the cases of $Tb_2Ti_2O_7$ (*45*) and $Tb_2Sn_2O_7$ (*46*).

The metallic nature of HoAgGe not only makes it a high-temperature (in comparison to pyrochlore spin ices) kagome ice, but may also lead to exotic phenomena, such as the interaction between electric currents and the magnetic monopoles or the toroidal moments, the relationship between the non-collinear ordering and the anomalous Hall effect (*47-50*), and metallic magnetoelectric effects caused by broken inversion symmetry (*51*). Our results suggest that ZrNiAl-type intermetallic compounds are a prototypical family of kagome spin systems, which may host other exotic phases beyond the classical spin liquid (*52, 53*) and deserve further investigation.

**Supplementary Materials**

Materials and Methods

Supplementary Text

Figs. S1 - S22

Table S1 - S4

References (55)-(65)

**Acknowledgement**

The authors would like to thank Yoshifumi Tokiwa, Satoru Nakatsuji, Oleg Tchernyshyov, Jianhui Xu, Junda Song, Viviane Peçanha-Antonio, and Martin Meven for helpful discussions, Anton Jesche for providing access to the MPMS, and Tim Delazzer for help with the CEF calculations. Kate A. Ross acknowledges Jeff Rau and Michel Gingras for CEF discussions. We are thankful to Klaus Kiefer and colleagues from Helmholz Zentrum Berlin for providing access and assistance using the VM4 magnet at MLZ. Technical assistance of Wolfgang Luberstetter for setting up the Oxford Instruments magnet on POLI is gratefully appreciated.

**Funding:** The work in Augsburg was supported by the German Science Foundation through SPP1666 (project number 220179758) and TRR80 (project number 107745057). The instrument POLI at Heinz Maier-Leibnitz Zentrum (MLZ), Garching, Germany, was operated by RWTH Aachen University in cooperation with FZ Jülich (Jülich Aachen Research Alliance JARA). Inelastic neutron experiments have been conducted at time-of-flight spectrometer NEAT operated by Helmholtz Zentrum Berlin. This work utilized the RMACC Summit supercomputer, which is supported by the National Science Foundation (awards ACI-1532235 and ACI-1532236), the University of Colorado Boulder, and Colorado State University. The Summit supercomputer is a joint effort of the University of Colorado Boulder and Colorado State University. The work in Prague was supported by the Czech Science Foundation through Project No. 18-10504S.

**Author contributions:** K. Z. and P. G. proposed the experiments; K. Z. synthesized single crystals, measured magnetic property and specific heat; H. D. and V. H. conducted single crystal elastic neutron scattering; V. P., H. D., and K. Z. refined magnetic structures; G. G. and M. R. conducted inelastic neutron scattering; H. C. provided theoretical analysis and MC simulation; K. A. R. performed CEF calculations; K. Z. , H. C. , and P. G. wrote the manuscript, with all authors shared their comments.

**Competing interests:** The authors declare that they have no competing interests.

**Data and materials availability:** The data presented in this paper can be found in (*54*).

**Table 1. Summary of HoAgGe single-crystal neutron data refinement results.**
Owing to the limited number of magnetic and nuclear peaks, and the uncertainty in aligning the field exactly with the *b* axis, the final refinement factors under field are usually larger than in the zero field case, yet still within a reasonable range (mostly smaller than 10%).

| Temperature(field) | 15K | 10K | 4K | 4K(1.5T) | 1.8K(2.5T) | 1.8K(4T) |
|---|---|---|---|---|---|---|
| (Magnetic)Space group | P-62m | P-6'm2' | P-6'm2' | Am'm2' | Am'm2' | Am'm2' |
| Magnetic vector: (k, k, 0) | | k=1/3 | k=1/3 | k=1/3 | k=1/3 | k=0 |
| Ho Label | Ho1 | Ho1-Ho3 | Ho1-Ho3 | Ho1-Ho6 | Ho1-Ho6 | Ho1, Ho2 |
| Ordered moment($\mu_B$) | | 5.2(1) | 7.5(1) | 7.6(1) | 7.6(2) | 7.4(3) |
| Neutron peaks & independent peaks | 535 & 106 | 330 & 99 | 971 & 217 | 234 & 164 | 254 & 157 | 220 & 137 |
| (Magnetic) Refinement factor: R,wR(%) | 2.95, 3.89 | 5.20, 6.27 | 3.38, 3.98 | 8.61, 10.85 | 5.90, 7.19 | 6.52, 8.18 |

**Table 2. The four low energy (< 1 meV) CEF modes of $Ho^{3+}$ in HoAgGe**

| Irreducible representation | Wave functions |
|---|---|
| $\Gamma_2$ | $-0.6186(\|7\!>+\|\!-\!7\!>) - 0.1871(\|5\!>+\|\!-\!5\!>) - 0.2591(\|3\!>+\|\!-\!3\!>) + 0.1234(\|1\!>+\|\!-\!1\!>)$ |
| $\Gamma_4$ | $-0.6209(\|7\!>-\|\!-\!7\!>) - 0.1961(\|5\!>-\|\!-\!5\!>) - 0.2636(\|3\!>-\|\!-\!3\!>) - 0.0814(\|1\!>-\|\!-\!1\!>)$ |
| $\Gamma_3$ | $-0.0780(\|8\!>-\|\!-\!8\!>) - 0.6256(\|6\!>-\|\!-\!6\!>) - 0.1512(\|4\!>-\|\!-\!4\!>) - 0.2822(\|2\!>-\|\!-\!2\!>)$ |
| $\Gamma_1$ | $0.0938(\|8\!>+\|\!-\!8\!>) + 0.6472(\|6\!>+\|\!-\!6\!>) + 0.1865(\|4\!>+\|\!-\!4\!>) + 0.1257(\|2\!>+\|\!-\!2\!>) - 0.2083\|0\!>$ |

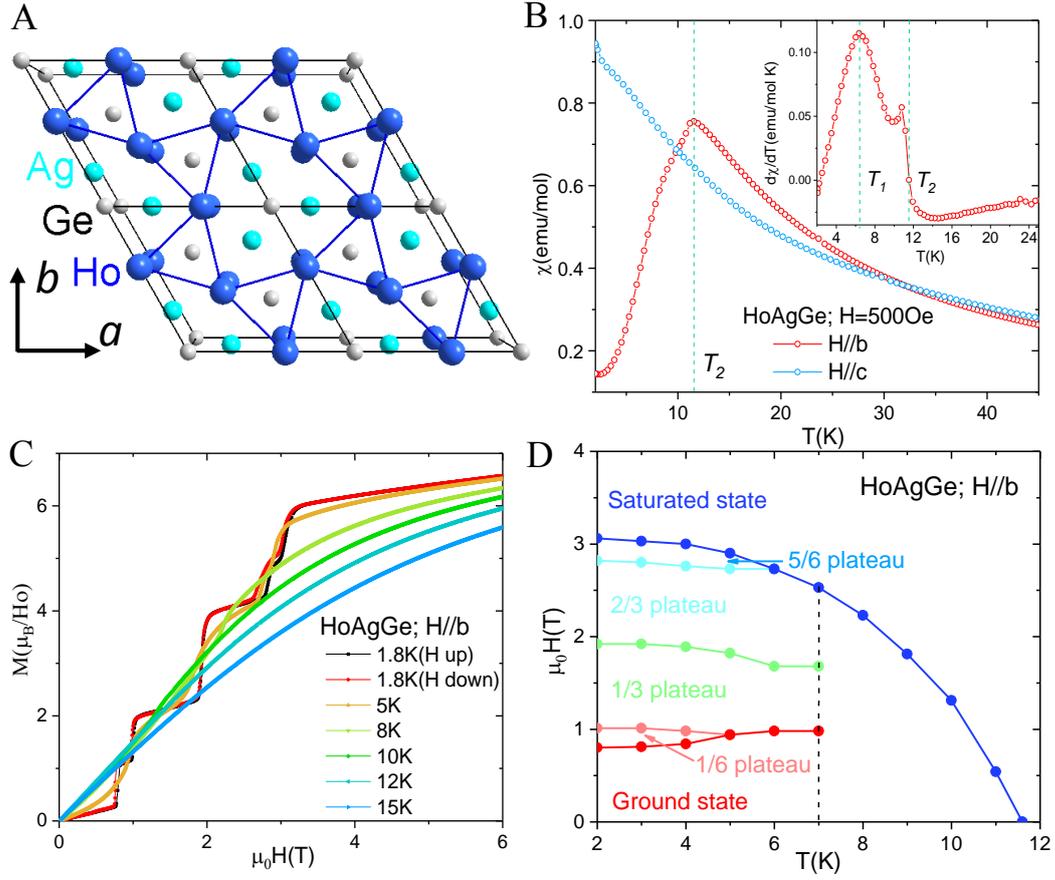

**Fig. 1: Crystal structure and magnetic properties of HoAgGe.** (A) *c*-axis projection of the HoAgGe crystal structure, with the definition of *a* and *b* directions. (B) Low-temperature susceptibility $\chi(T)$ of HoAgGe for both **H**//*b* and **H**//*c* under 500 Oe, with $d\chi(T)/dT$ in the inset. (C) Isothermal in-plane (**H**//*b*) magnetization for HoAgGe at various temperatures. (D) Summarizes the dependence of the metamagnetic transitions on temperature, with the dotted line indicating $T_1$ (see text).

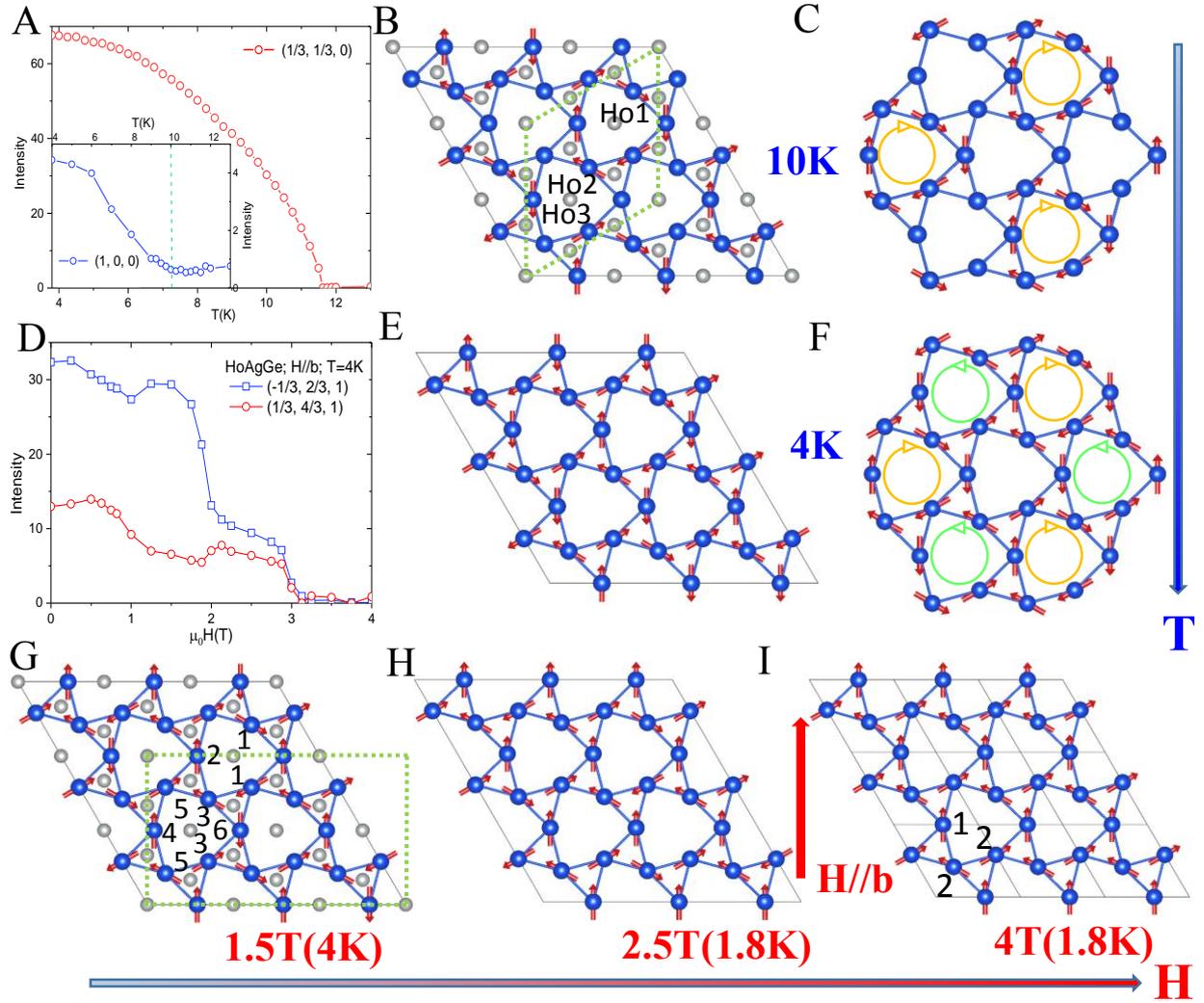

**Fig. 2: Magnetic structures of HoAgGe versus temperature and field with H//*b*.** (A) the neutron diffraction intensity of the magnetic peak (1/3, 1/3, 0) from 13K down to 3.8K, with the intensity of the nuclear site (1, 0, 0) as an inset. (B) the refined magnetic structures of HoAgGe at 10K. The magnetic unit cell is indicated by the green rhombus, with the three inequivalent Ho sites labeled by Ho1, Ho2, and Ho3. (C) The counterclockwise hexagons of spins in the partially ordered structure of HoAgGe at 10K, with one third spins not participating in the long-range order. (D) Intensity of the magnetic peak (-1/3, 2/3, 1) and (1/3, 4/3, 1) versus field at 4K. (E) The refined magnetic structure of HoAgGe at 4K. (F) The clockwise and counterclockwise hexagons of spins in the magnetic structure of HoAgGe at 4K, which is exactly the expected $\sqrt{3} \times \sqrt{3}$ ground state of kagome spin ice. (G) The refined magnetic structure of HoAgGe at *H*=1.5T and *T*=4K. The refinement was done in the $3 \times \sqrt{3}$ light green rectangle. The six inequivalent Ho sites are labeled by number 1-6 for simplicity. (H) The refined magnetic structure of HoAgGe at *H*=2.5T and *T*=1.8K. (I) The refined magnetic structure of HoAgGe at *H*=4T and *T*=1.8K, with the two inequivalent Ho sites labeled by 1 and 2. The field direction is marked by the red arrow for G-I.

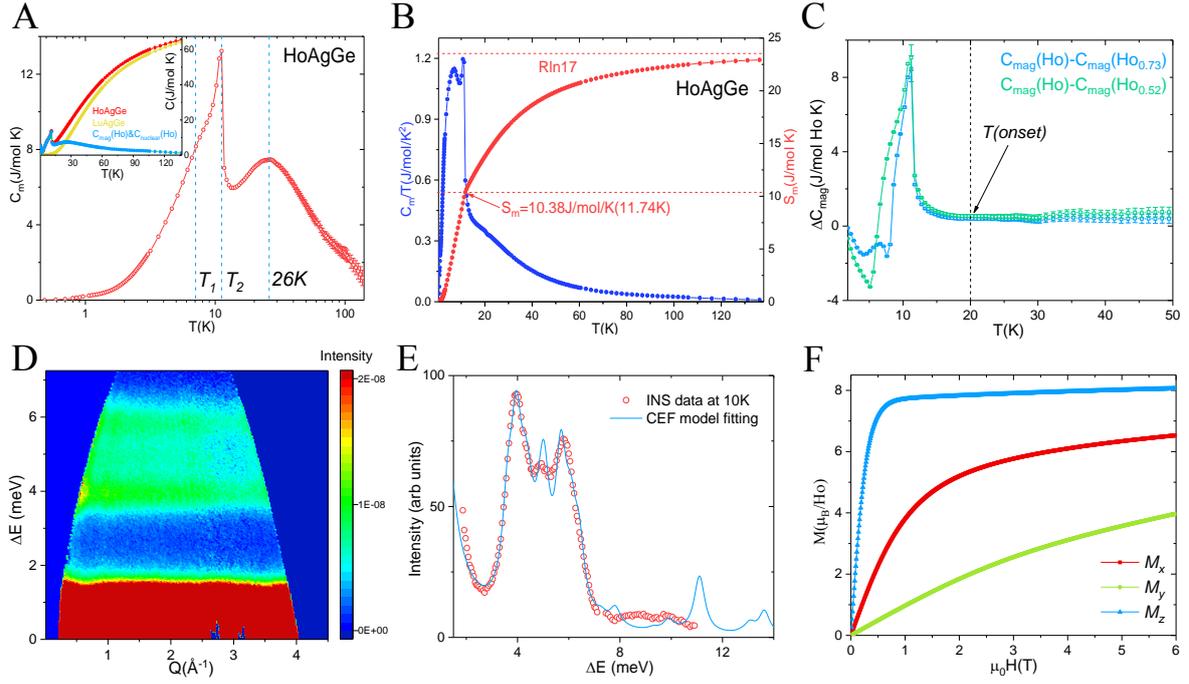

**Fig. 3: Magnetic specific heat and INS results of HoAgGe.** (A) Magnetic contribution to the specific heat $C_m$ of HoAgGe with the dotted lines indicating $T_1$, $T_2$, and broad peak at 26K (see text). Note that the error bars below 30 K are smaller than the symbol sizes. Inset: Specific heat of HoAgGe, LuAgGe, and their difference. The latter is defined as the sum of the magnetic and the nuclear contributions to the specific heat of HoAgGe. (B) $C_m/T$ data and the corresponding magnetic entropy $S_m$, which tends to saturate at the theoretical value of Rln17 above 100K. (C) Difference between the magnetic specific heat of HoAgGe and that of $Lu_{1-x}Ho_xAgGe$ ($x$=0.52 and 0.73) after normalization (see text). (D) INS spectra of HoAgGe at 10K with incident neutron wavelength 3Å. (E) Constant-Q cuts (1.4 < Q < 2.2 Å$^{-1}$) showing the results of the CEF fitting to neutron scattering data. (F) Isothermal magnetization calculated for CEF fitting parameters at 1.5K for three quantization axes.

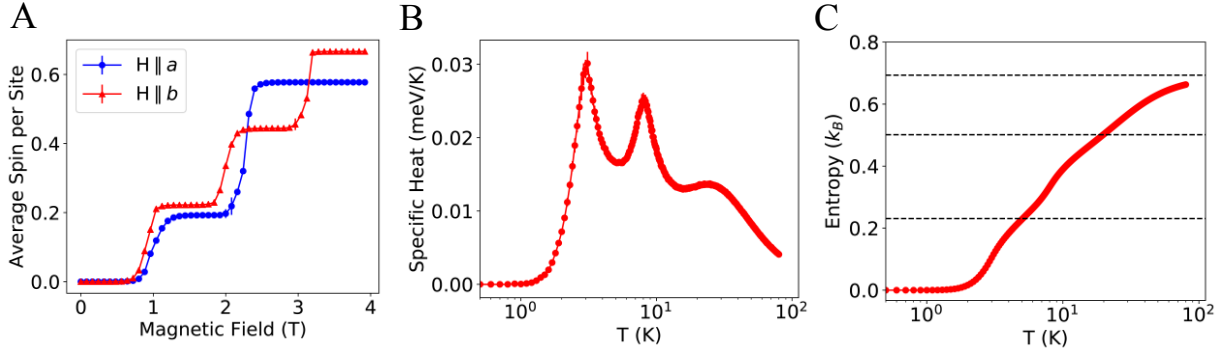

**Fig. 4: Monte Carlo simulations of the 2D classical spin model for HoAgGe.** (A) *M(H)* curves at 1K for **H** along *a* and *b* axes, respectively. (B) Temperature dependence of the specific heat per spin. (C) Magnetic entropy per spin calculated from the specific heat. The three horizontal dashed lines correspond to $\ln 2 \approx 0.693$ (paramagnetic Ising), 0.501 (short-range ice order), and $\frac{1}{3}\ln 2 \approx 0.231$ (toroidal order), respectively. An 18×18 cell is used for the calculation.